\documentclass[prb,aps,twocolumn]{revtex4}
\usepackage{amsmath}
\usepackage{amssymb}
\usepackage{bm}
\usepackage{epsfig}
\usepackage{graphicx}
\begin{document}

\title{Fine structure of exciton excited levels in a quantum
dot with a magnetic ion}

\author{M.M.~Glazov, E.L.~Ivchenko}
     \address{A.F. Ioffe Physico-Technical Institute,
     Russian Academy of
Sciences, 194021 St.-Petersburg, Russia} 

\author{L.~Besombes, Y.~L\'eger, L.~Maingault and H.~Mariette}
     \address{CEA-CNRS group ``Nanophysique et Semiconducteurs'',
     Laboratoire de spectrom\'{e}trie physique,
CNRS and Universit\'{e} Joseph Fourier-Grenoble 1, BP 87, F-38402 St
Martin d'H$\grave{e}$res, France} 

\begin{abstract}
The fine structure of excited excitonic states in a quantum dot
with an embedded magnetic ion is studied theoretically and
experimentally. The developed theory takes into account the
Coulomb interaction between charged carriers, the anisotropic
long-range electron-hole exchange interaction in the
zero-dimensional exciton, and the exchange interaction of the
electron and the hole with the $d$-electrons of a Mn ion inserted
inside the dot. Depending on the relation between the quantum dot
anisotropy and the exciton-Mn coupling the photoluminescence
excitation spectrum has a qualitatively different behavior. It
provides a deep insight into the spin structure of the excited
excitonic states.
\end{abstract}

\maketitle                   

\section{Introduction}
Optical microspectroscopy is a set of powerful optical techniques
that do not damage the sample and allow one to extract local
quantitative information on micron and submicron scales. These
high spatial resolution optical techniques has been used to
measure the photoluminescence (PL) from single quantum dots and
study the exchange interaction between a zero-dimensional exciton
and a few \cite{bacher02,mackowski05} or a single magnetic ion
embedded in the quantum dots
(QD)~\cite{besombes04,besombes05,leger05,leger05bis,leger06}.
Theoretically, the ground state energy spectrum of a quantum dot
containing a Mn ion was studied in
Refs.~\onlinecite{govorov,rossier,qu06}.

In this work we have theoretically and experimentally studied the
fine structure of \emph{excited} excitonic states in such
magnetically-doped QDs. We have developed a theory taking into
account both direct Coulomb and long-range exchange interaction
between an electron and a hole in the zero-dimensional exciton and
exchange interaction of the coupled electron-hole pair with the Mn
$d$-electrons. The fine structure of the excited state is shown to
be determined by an interplay of the anisotropic electron-hole
long-range exchange interaction, Coulomb interaction between the
charge carriers and the coupling of an exciton with a Mn ion.

The developed theory is compared with the PL excitation (PLE)
measurements on an individual QD carrying a Mn$^{2+}$ ion. The
coupling constants of exciton-Mn interaction for the ground and
optically active excited states are determined. The analysis of the
fine structure of the excited states allows us to determine the
degree of QD anisotropy and make conclusions about the magnetic ion
position in a quantum dot.

The paper is organized in a following way: Sec. II is devoted to
the theoretical analysis of the fine structure of $pp$-shell
excitonic states in the QD without a Mn ion, Sec. III presents the
results on the exciton interaction with the magnetic ion, and Sec.
IV presents the experimental results and their analysis.

\section{Exciton states in non-magnetic dots}
We consider heavy-hole exciton states in a single quantum-well
quantum dot. This means that we neglect heavy-light hole mixing
and assume an exciton to be formed by a conduction-band electron
with the spin $s = \pm 1/2$ and a heavy hole with the spin (or
angular-momentum component) $j = \pm 3/2$.~\cite{leger05bis} For
the in-plane confinement of free carriers, we choose {\it coaxial}
{\it parabolic} potentials with $z$ as the principal axis. In what
follows we concentrate on QDs of sizes small enough to have the
in-plane localization lengths for electrons ($a_e$) or holes
($a_h$) smaller than the two-dimensional (2D) exciton Bohr radius
$a_{\mathrm B}$.

To begin with, let us consider an axially (cylindrically) symmetric
QD where the exciton ground state ($ss$-shell) is described by the
envelope
\begin{equation}\label{s}
S(\rho_e, \rho_h) = \frac{1}{\pi a_e a_h} \exp{ \left(
-\frac{\rho_e^2}{2 a^2_e} - \frac{\rho_h^2}{2 a^2_h} \right) }\ ,
\end{equation}
where ${\bm \rho}_{e,h}$ is the electron or hole in-plane radius
vector with the components $x_e,y_e$ or $x_h,y_h$ referred to the
center of the confining potential. Similarly, the $pp$-orbital
excited states being the products of electron and hole $P_{x,y}$
orbitals have the form
\begin{widetext}
\begin{equation} \label{dxy}
D_{xx}(\bm \rho_e,\bm \rho_h) = 2\ \frac{x_e x_h}{a_e a_h} \
S(\rho_e, \rho_h) \ , \ D_{yy}(\bm \rho_e,\bm \rho_h) = 2\
\frac{y_e y_h}{a_e a_h} \ S(\rho_e, \rho_h) \ ,
\end{equation}
\[
D_{xy}(\bm \rho_e,\bm \rho_h) = 2\ \frac{x_e y_h}{a_e a_h} \
S(\rho_e, \rho_h)\ , \ D_{yx}(\bm \rho_e,\bm \rho_h) = 2\
\frac{y_e x_h}{a_e a_h} \ S(\rho_e, \rho_h)\ ,
\]
\end{widetext}
where the subscripts in $D_{xx}, D_{yy}, D_{xy}$, and $D_{yx}$
describe the symmetry of the $P$-orbitals for an electron and a
hole.

\begin{figure*}[htb]
\includegraphics[width=0.95\linewidth]{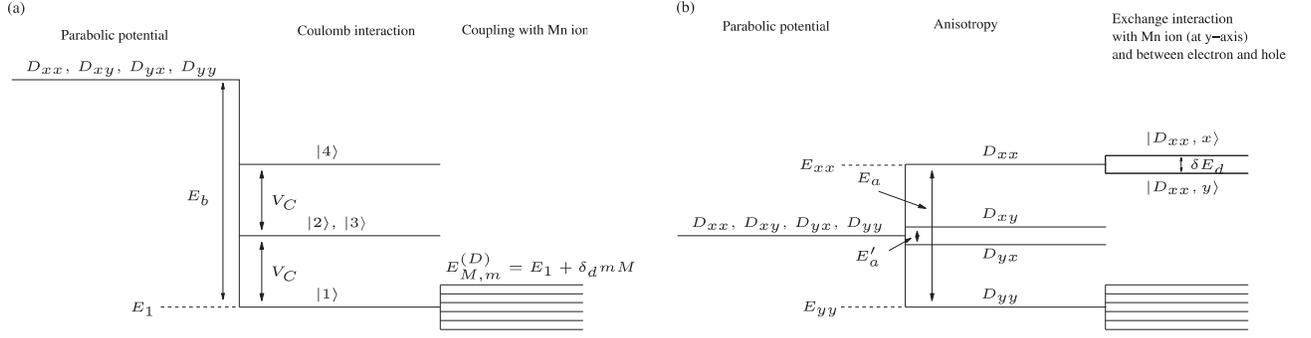}
  \caption{Schematic $pp$-shell energy-level diagram
  for the axial (a) and anisotropic QD (b). The energy splittings
  are shown not to scale.
  }\label{fig:levels1}
\end{figure*}

In the isotropic parabolic potential the four $pp$-shell states
are degenerate. The Coulomb interaction between an electron and a
hole lifts this degeneracy, see Fig.~\ref{fig:levels1}a. The
lowest lying state with the wavefunction
\begin{equation} \label{single}
|1\rangle = \frac{1}{\sqrt{2}}\ [D_{xx}(\bm \rho_e,\bm \rho_h) +
D_{yy}(\bm \rho_e,\bm \rho_h)]
\end{equation}
is optically active. The states $|2\rangle = [D_{xx}(\bm
\rho_e,\bm \rho_h) - D_{yy}(\bm \rho_e,\bm \rho_h)]\sqrt{2}$ and
$|3\rangle = [D_{xy}(\bm \rho_e,\bm \rho_h) + D_{yx}(\bm
\rho_e,\bm \rho_h)]/\sqrt{2}$ are degenerate and they are shifted
above the state $|1\rangle$ by the energy
\begin{equation}\label{Vc}
V_C = \frac{3\sqrt {\pi }}{4} \frac {\zeta^{2} }{
(1+\zeta^2)^{5/2} }\ \frac{e^2}{\varkappa_0 a_e}\ ,
\end{equation}
where $\varkappa_0$ is the low-frequency dielectric constant and
$\zeta = a_h/a_e$. The splitting between the pair of states
$|2\rangle$, $|3\rangle$ and the highest state $|4\rangle
=[D_{xy}(\bm \rho_e,\bm \rho_h) - D_{yx}(\bm \rho_e,\bm
\rho_h)]/\sqrt{2}$ is given by $V_C$ as well. The binding energy
of the exciton $| 1 \rangle$ referred to the unperturbed position
of $pp$-orbitals reads
\begin{equation}\label{Eb}
E_b  = \frac{\sqrt{\pi} }{2} \frac
{\zeta^4+5\zeta^2+1}{(1+\zeta^2)^{5/2}}\ \frac{e^2}{\varkappa_0
a_e} \:.
\end{equation}
The parameters $E_b$ and $V_C$ completely determine the direct
Coulomb interaction in the parabolic QD.

We next turn to the slightly elliptical QD where the effective
localization radii $a_e$ and $a_h$ are replaced by localization
lengths $a_{e,\hspace{0.5 mm} i}, a_{h,\hspace{0.5 mm} i}$ ($i =
x, y$) in the $x$ and $y$ directions. The anisotropy is
characterized by the ratios
$$
\beta_e = \frac{ a_{e,\hspace{0.5 mm} y} - a_{e, \hspace{0.5 mm}
x}}{2a_e}\:,\: \beta_h = \frac{a_{h,\hspace{0.5 mm} y} - a_{h,
\hspace{0.5 mm} x}}{2a_h}\:.
$$
For simplicity we assume $\beta_e = \beta_h \equiv \beta$ and
$\beta \ll 1$. The anisotropy fully lifts the degeneracy of
$pp$-orbital states. In the case of large anisotropy (the
criterion is given below) the splitting between two bright states
$D_{xx}(\bm \rho_e, \bm \rho_h)$ and $D_{yy}(\bm \rho_e, \bm
\rho_h)$  is given (to the lowest order in $\beta$) by
\begin{equation}\label{anisosplit}
E_a = E_{xx} - E_{yy} = \left(1+ \frac{\sigma}{\zeta^2}\right)\
\frac{4\hbar^2 \beta}{m_e a_e^2}\  ,
\end{equation}
where $\sigma=m_e/m_h$ characterizes electron-hole mass ratio. The
splitting between the dark states, $D_{xy}(\bm \rho_e, \bm
\rho_h)$ and $D_{yx}(\bm \rho_e, \bm \rho_h)$, arises due to the
difference between carrier masses and/or localization radii,
namely,
\begin{equation}\label{anisosplit1}
E_a' = E_{xy} - E_{yx} = \left(1 - \frac{\sigma}{\zeta^2}\right)\
\frac{4\hbar^2 \beta}{m_e a_e^2}\ .
\end{equation}
The transition between the Coulomb-dominated and
anisotropy-dominated regimes takes place at $|E_a| \sim V_C$.

The long-range exchange interaction affecting the bright exciton
with $m = \pm 1$ is described by~\cite{efros98,glazov05}
\begin{equation}\label{Hlt_gen}
\mathcal H^{(\rm long)}_{n'n}  = \frac{1}{4\pi\varkappa_\infty}
\left(\frac{e\hbar |p_0|}{m_0 E_g} \right)^2 \int d\bm K
\frac{K^*_{m'} K_m }{K}\ \widetilde{\psi}^{*}_{n'}(\bm K)
\widetilde{\psi}_n(\bm K),
\end{equation}
where $n$ (or $n'$) is the exciton-state index including both the
angular-momentum component $m$ (or $m'$) and orbital state,
$\varkappa_\infty$ is the high-frequency dielectric constant,
${\bm K}$ is the two-dimensional center of mass wave vector with
the components $K_x, K_y$, $K_{\pm 1} = K_x \pm {\rm i} K_y$,
$m_0$ is the free electron mass, $E_g$ is the band gap, $p_0$ is
the interband matrix element of the momentum operator, and we
introduced the 2D Fourier-transform $ \widetilde{\psi}(\bm K) =
\int d{\bm \rho}\: e^{-{\rm i} \bm K {\bm \rho}} \Psi({\bm \rho},
{\bm \rho})$ of the exciton envelope function $\Psi(\bm \rho_e,\bm
\rho_h)$ taken at the coinciding coordinates, $\bm \rho_e = \bm
\rho_h \equiv {\bm \rho}$. For the $ss$-shell, $\Psi_n(\bm
\rho,\bm \rho) = S(\bm \rho,\bm \rho)$ and, for the $pp$-shell
states $| D_{\alpha \beta}, m \rangle$, $\Psi_n(\bm \rho,\bm \rho)
= D_{\alpha \beta}(\bm \rho,\bm \rho)$ with $\alpha$ and $\beta$
running through $x,y$.

In the isotropic QD the $ss$-shell radiative doublet is not split
by the interaction long-range interaction Eq. \eqref{Hlt_gen}. We
introduce the characteristic value of the long-range exchange for
$pp$-shell states
\begin{equation}\label{elt}
\mathcal E = \left( \frac{ e \hbar |p_0|}{m_0 E_g} \right)^2
\frac{\sqrt{\pi}}{16 \varkappa_\infty a_e^3 \zeta\sqrt{1+\zeta^2}
}\:.
\end{equation}
For CdTe QD with $a_e = 45$~\AA, $a_h = 90$~\AA, $\varkappa_0 =
10.4$, $\varkappa_\infty = 7.1$, $2p_0^2/m_0 = 17.9$~eV, $E_g =
1.6$~eV [see Ref. \onlinecite{efros98}] the Coulomb interaction
induced splitting $V_C\sim 3$~meV and $\mathcal E\sim 1.5\times
10^{-2}$~meV. Thus, $\mathcal E\ll V_C$ and the exchange
interaction should be considered as a correction to the direct
Coulomb interaction.~\cite{fn1} Therefore, in an axial QD the
exchange-induced splitting of the radiative $pp$-orbital doublet
is proportional to $\mathcal E^2/V_C$ and can be neglected.

If the QD possesses an anisotropy in the axes $x,y$ the radiative
doublet of the ground state (\ref{s}) is split into a pair of
linearly polarized sublevels with the microscopic dipole momentum
parallel to $x$ and $y$. This anisotropic splitting is given by
\begin{equation}\label{deltaEs}
\delta E_s = 24 \beta \mathcal E\ ,
\end{equation}
and is proportional to the anisotropy degree $\beta$ entering also
into Eqs.~(\ref{anisosplit}), (\ref{anisosplit1}). If anisotropic
splitting, $E_a$, of the $pp$-shells $D_{xx}$, $D_{yy}$ exceeds
the Coulomb energy $V_C$, each shell is additionally split into
radiative sublevels $|D_{xx},x\rangle, {|D_{xx},y\rangle}$ or
$|D_{yy}, x\rangle, {|D_{yy}, y\rangle}$ according to
\begin{equation}\label{deltaEd}
\delta E_d = E_{|D_{xx},x\rangle} - E_{|D_{xx},y\rangle} =
E_{|D_{yy},y\rangle} - E_{|D_{yy},x\rangle} = 3\mathcal E\: .
\end{equation}
Comparing Eqs.~(\ref{deltaEs}) and (\ref{deltaEd}) we conclude
that splittings of the $ss$- and $pp$-shells differ by $8 \beta$.

\section{Interplay between exciton-M{\lowercase{n}} and electron-hole
exchange interaction} The Hamiltonian describing exciton-Mn
exchange interaction can be conveniently represented in the basis
of the certain components $s,j,M$ of the electron, hole and Mn-ion
spins as
\begin{equation}\label{hamz}
 \mathcal H_{\rm
Mn} = [ A_e s\ \delta(\bm r_0 - \bm r_e) + A_h j\ \delta(\bm r_0 -
\bm r_h)] M\:,
\end{equation}
where $\bm r_0$ is the Mn ion position, $A_e$ and $A_h$ are the
coupling parameters for conduction and valence band
electrons.~\cite{merkulov99} In derivation of Eq. \eqref{hamz} we
took into account only $j=\pm 3/2$ hole states and neglected the
interaction with the electron in-plane spin since, for the
realistic set of parameters, $A_e < A_h/2$, see
Refs.~\onlinecite{besombes04,rossier}.

In a QD with an embedded Mn ion, fine structure of the
ground-state exciton is determined by the interplay between the QD
anisotropy and the exchange coupling with the Mn ion. From now on
we consider the most important (from both experimental and
theoretical points of view) case where the ground state
anisotropic splitting Eq.~\eqref{deltaEs} is negligible as
compared with the coupling with a magnetic ion. Without the
exciton-ion exchange interaction, the ground state of the system
``exciton + Mn ion'' is 4$\times$6-fold degenerate. The
interaction $\mathcal H_{\mathrm{ Mn}}$ splits this state into 12
doubly-degenerate sublevels half of which, namely, those with the
exciton angular-momentum component $m \equiv s+j = \pm 1$, are
optically active (or bright). Their energies are given by
\begin{equation} \label{sublevelsa}
E_{m,M}^{(S)} = E_0^{(S)} + \delta_s m M\:,\:\delta_s
=(3I_h^{(s)}-I_e^{(s)})/2\:,
\end{equation}
where $E_0^{(S)}$ is the $ss$-shell exciton energy,
$$I_e^{(s)} =
\frac{A_e}{\pi a^2_e} \mathrm {e}^{-\rho_0^2/a_e^2}
\varphi^2_{e1}(z_0)\:,I_h^{(s)} = \frac{A_h}{\pi a^2_h} \mathrm
e^{-\rho_0^2/a_h^2} \varphi^2_{hh1}(z_0)\:,$$ $\rho_0^2 = x^2_0 +
y^2_0$; $x_0, y_0$ and $z_0$ are the Cartesian components of ${\bm
r}_0$ and the envelope $\varphi_{e1}(z)$ or $\varphi_{hh1}(z)$
describes the electron or hole confinement along $z$.

The fine structure of the $pp$-shell exciton excited states is
determined by a combined effect of the exciton-Mn and the
electron-hole (long-range) exchange interactions. If the QD
anisotropy is small so that $|E_a| \ll V_C$ the only one bright
state $| 1 \rangle$ defined by Eq.~(\ref{single}) splits into $6$
doubly-degenerate sublevels, see Fig.~1a,
\begin{equation} \label{sublevelsD}
E_{m,M}^{(D)} = E_1 + \delta_d m M\:,
\end{equation}
where $E_1$ is the energy of $|1\rangle$ state,
\begin{equation}\label{deltaD}
\delta_d ={1\over 2} \bigl(3I_h^{(d)}-I_e^{(d)}\bigr),
\end{equation}
and the coupling constants
\begin{equation}\label{Id}
I_e^{(d)} = \frac{r_0^2}{a_e^2}\ I_e^{(s)}\ ,\: I_h^{(d)} =
\frac{r_0^2}{a_h^2}\ I_h^{(s)}\ ,
\end{equation}
are strongly sensitive to the position of the Mn ion. For example,
if the Mn ion is located exactly in the QD center the splitting of
$pp$-shell states vanishes.

In highly anisotropic QDs where $V_C\ll E_a$ the fine structure of
the $D_{xx}$ and $D_{yy}$ levels is even more rich. In general
case an overlap between exciton and Mn ion is different for these
two orbitals and one needs to introduce two coupling constants for
each type of the carriers, namely,
\begin{equation}\label{Idxxyy}
I_e^{(xx)} = \frac{x_0^2}{a_e^2}\ I_e^{(s)}\ ,\: I_h^{(xx)} =
\frac{x_0^2}{a_h^2}\ I_h^{(s)}\ ,
\end{equation}
\[
I_e^{(yy)} = \frac{y_0^2}{a_e^2}\ I_e^{(s)}\ ,\: I_h^{(yy)} =
\frac{y_0^2}{a_h^2}\ I_h^{(s)}\ .
\]
The eigen energies are determined by the interplay of the
exciton-Mn interaction described by the parameter
\begin{equation}\label{deltaXXYY}
\delta_{ii} =3 I_h^{(ii)}- I_e^{(ii)}\hspace{ 8 mm} (i=x,y)\:,
\end{equation}
and the long-range exchange splitting Eq.~\eqref{deltaEd}. For a
fixed orbital $D_{ii}$ one should observe six doubly-degenerated
lines with energies
\begin{equation}\label{interplay}
E^{(ii)}_{m,M} = E_{ii} \pm \sqrt{\left(\frac{\delta
E_d}{2}\right)^2 + \left(\delta_{ii} M\right)^2 },
\end{equation}
where $E_{ii}$ is the position of the corresponding unperturbed
orbital. For $E_d \neq 0$, the levels are nonequidistant and the
crossover between anisotropy dominated regime with two linearly
polarized states ($\delta E_d \gg \delta_{ii}$) and the regime of
dominant exciton-Mn interaction ($\delta E_d \ll \delta_{ii}$)
with six equidistant sublevels takes place.

\begin{figure}[htbp]
\includegraphics[width=1.05\linewidth]{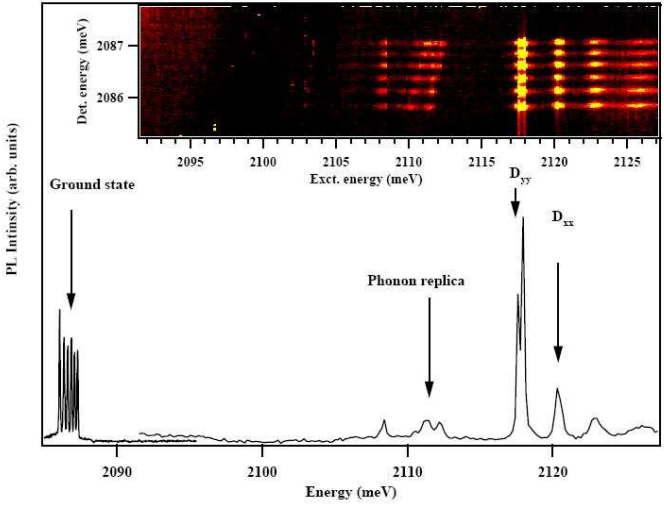}
\caption{(Color online) Experimental PL and PLE spectra of Mn-doped
quantum dot exciton. The spectral features related to the ground and
excited states are marked correspondingly. The inset shows the
contour plot of the multi-channel PLE.}\label{fig:exper1}
\end{figure}

\section{Discussion and comparison with an experiment}

We use micro-spectroscopy to analyze the optical properties of
individual Mn-doped self-assembled CdTe/ZnTe QDs. Single Mn atoms
are introduced in CdTe/ZnTe QDs
\cite{besombes04,besombes05,leger05} by adjusting, during the
growth process, the density of Mn atoms to be roughly equal to the
density of QDs.~\cite{Maingault06} The PL of individual QDs is
excited with a tunable dye laser and collected through aluminium
shadow masks with $0.2 \ldots 1.0$ $\mu$m apertures. The PL is
then dispersed by a 2-m additive double monochromator and detected
by a nitrogen cooled Si charge-coupled device (CCD).

\begin{figure}[htbp]
\includegraphics[width=0.925\linewidth]{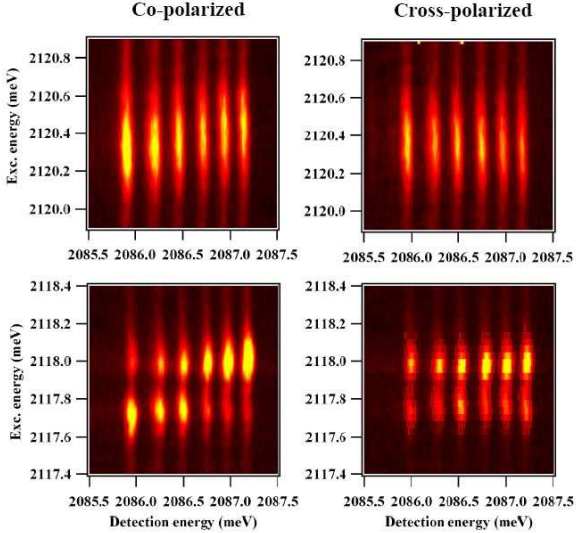}
\caption{(Color online) PLE contour plots for excited states
D$_{xx}$ (upper panel) and D$_{yy}$ (lower panels) obtained for
co-polarized (left panels) and cross-polarized (right panels)
circular excitation.}\label{fig:exper2}
\end{figure}

\begin{figure}[htbp]
\includegraphics[width=0.825\linewidth]{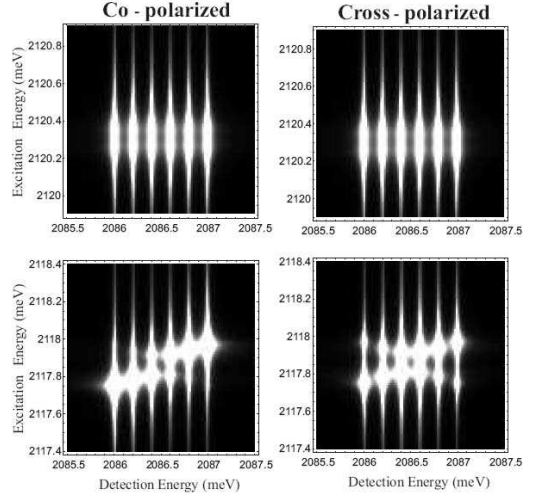}
\caption{Calculated PLE spectra in co-polarized (left) and
cross-polarized (right) configuration. The parameters used are as
follows: $E_{yy} = 2117.86$~meV, $E_{xx} = 2120.30$~meV, $\delta
E_{d} = 0.05$~meV, $\tilde \delta_{xx}=0.01$~meV, $\tilde
\delta_{yy} = 0.04$~meV, $\delta_s=0.2$~meV and $E_0^{(S)} =
2086.5$~meV. The lines are Lorentzian-broadened with the widths
$\Gamma_s=0.03$~meV and $\Gamma_{xx}=0.06$~meV,
$\Gamma_{yy}=0.03$~meV. }\label{fig:theorCirc}
\end{figure}

\begin{figure}[htbp]
\includegraphics[width=0.925\linewidth]{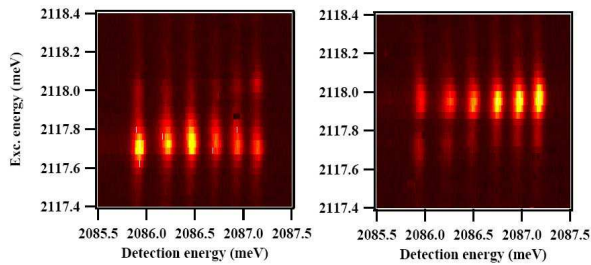}
\caption{(Color online) PLE spectra of D$_{yy}$ states for linearly
polarized excitation and unpolarized detection. Left and right
panels correspond to two orthogonal linearly polarized excitations.
}\label{fig:experLin}
\end{figure}

The experimental findings for a Mn-doped QD are summarized in
Figs.~\ref{fig:exper1}, \ref{fig:exper2}, and \ref{fig:experLin}.
This particular QD shows clear linearly polarized excited states.
The analysis of the fine structure of other QDs presenting linearly
polarized excited states leads qualitatively  to the same
conclusions. We also note that some QDs demonstrating the excited
states in the same energy range present a very weak or
non-detectable linear polarization splitting. Fig.~\ref{fig:exper1}
shows PL and PLE spectra of this QD. The ground state PL
demonstrates 6 equidistant lines positioned at $\sim 2086$~meV.
These lines correspond to different projections of Mn spin (see Eq.
\eqref{sublevelsa}). Their approximately equal intensities and
regular energy spacing evidence the negligible anisotropic splitting
of the ground state. The splitting between these lines suggests the
value $\delta_s\approx 0.2$~meV being in a good agreement with
previous studies.~\cite{besombes04}

The first absorption lines in the measured PLE
(Fig.~\ref{fig:exper1}) lie by $\sim 25$~meV above the ground state.
This energy range corresponds to the LO-phonon energy between CdTe
and ZnTe, thus this feature can be identified with the phonon
replica of the ground state. The non-trivial PLE pattern coming from
the replica is a consequence of complicated phonon spectra in such
QDs system.

Now we proceed with the discussion of the excited-states fine
structure. Two excited states with energies $\approx 2117.86$~meV
and $2120.30$~meV are seen in Fig.~\ref{fig:exper1}. We relate
them with the $D_{yy}$- and $D_{xx}$-orbital states. The
difference in their intensities can be attributed to (i) the
shorter lifetime of the highest energy state and (ii) the Coulomb
interaction which, according to Eqs.~\eqref{single} and
\eqref{Vc}, intermixes the $D_{xx}$ and $D_{yy}$ states. The
latter effect reduces the oscillator strength of the higher state
and, as a result, the anisotropic exchange splitting of this
state.

A value of the ratio between the energy separation of the excited
states $D_{xx}$ and $D_{yy}$ and the distance between ground and
excited states allows one to estimate from Eq. \eqref{anisosplit}
the quantum dot anisotropy degree $\beta$ to be $\approx 0.02$
which supports the assumption that the anisotropic splitting of
$ss$-shell state is not observed. Figure~\ref{fig:exper1} clearly
manifests the splitting of $D_{yy}$ state into a pair of lines. We
assume this splitting to come from the QD anisotropy [see Eq.
\eqref{deltaEd}]. The splitting of $D_{xx}$ state is not observed
for the reasons suggested above.

The comparison between the experimental data and theoretical
predictions in Figs.~\ref{fig:exper2} and \ref{fig:theorCirc},
respectively, shows that the best possible agreement of the data
is obtained for anisotropic splitting $\delta E_{d} = 0.05$~meV
and $\tilde \delta_{yy} = 0.04$~meV. The order of magnitude of
$\delta E_{d}$ is consistent with the quantum dot size $a_e =
45$~\AA~and $a_h = 90$~\AA~(i.e. $\zeta=2$). It agrees with other
results for II-VI QDs where the hole confinement is weaker than
that of electrons.~\cite{zhang97} For these parameters the
predicted energy separation between the ground and excited states
is $43$~meV which is somewhat larger than the experimentally
observed $33$~meV.

Fully quantitative description of the experimental results would
require a more complicated theory taking into account all details
of the band structure and confinement potential as well as the
short-range contribution to the exchange
interaction.~\cite{aleiner, franceschetti}

\begin{figure}[htbp]
\includegraphics[width=0.825\linewidth]{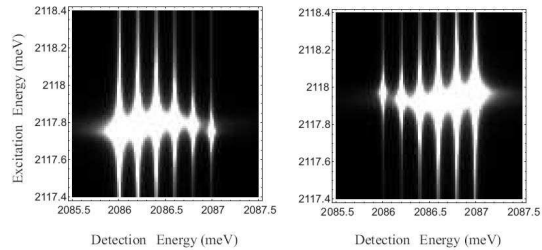}
\caption{Calculated PLE spectra for linearly polarized excitation
and unpolarized detection. Left and right panels correspond to two
orthogonal linear polarizations. The parameters used in the
calculation are given in the caption of Fig.~\ref{fig:theorCirc}
}\label{fig:theorLin}
\end{figure}

In order to have a deeper insight into the fine structure of the
$D_{yy}$ state we have performed the PLE measurements for the
linearly polarized excitation and unpolarized detection. The
experimental results are given in Fig.~\ref{fig:experLin}. The
theoretical PLE plots calculated for the same parameters as the
spectra for circularly-polarized excitation are shown in
Fig.~\ref{fig:theorLin}. The reasonable agreement between the
theory and experiment is seen. We note that we do not introduce
any spin relaxation in the calculations. The similarity in the PL
intensity distribution obtained in the theory and in the
experiment shows that under resonant excitation the spin
relaxation time of the exciton-Mn complex is much more longer than
the lifetime of the exciton: no significant relaxation occurs
during the exciton lifetime. As shown in Figs.~\ref{fig:exper2}
and~\ref{fig:experLin}, this long spin relaxation time combined
with the fine structure of the excited states permits to create
selectively a given spin configuration of the exciton-Mn complex
by tuning the polarization and wavelength of the excitation laser.

Let us note finally that the experimentally observed PLE intensity
distribution for the higher energy state $D_{xx}$ is almost
uniform, therefore we conclude that the exciton-Mn ion coupling is
smaller for this state as compared to $D_{yy}$. We can thus
conclude that the Mn ion is located nearby the $y$ axis. If we
completely neglect electron-Mn interaction, using Eq.
\eqref{deltaXXYY} and the fitted value $\tilde \delta_{yy} =
0.04$~meV we deduce that the Mn-ion is positioned at $y_0\approx
0.3 a_h$, $x_0 \ll y_0$.

\section{Conclusion}

In conclusion, we have performed a combined experimental and
theoretical study of the fine structure of exciton excited states
in the QDs containing a single magnetic ion. We have identified
two regimes of uniaxial and anisotropic quantum dots where the
qualitatively different level arrangement and PLE spectra are
predicted. The theoretical results are compared with the
experimental data on photoluminescence and photoluminescence
excitation for an individual Mn-doped QD. All important
experimental observations are reproduced theoretically. The
comparison has made possible to determine the Mn-exciton
interaction constants, QD anisotropy degree and to estimate the
position of the Mn ion in the quantum dot.

\begin{acknowledgements}
  The work was partially supported by RFBR, ``Dynasty'' foundation
-- ICFPM and French ANR contract MOMES.
\end{acknowledgements}



\begin{thebibliography}{10}
\bibitem{fn1} We note that for $p$-shell excited
states the Coulomb interaction is the same for $P_x$ and $P_y$
states and therefore, in the axial approximation, and the fine
sublevels of $p$-shell are split due to long-range exchange only,
see Ref.~\onlinecite{glazov05}
\end{thebibliography}
\end{document}